\documentclass[twocolumn,english,aps,floatfix,superscriptaddress,showpacs,amsfonts,amssymb,superscriptaddress]{revtex4-1}
\usepackage{color}
\usepackage{graphicx}
\usepackage{amsmath}
\usepackage{latexsym}
\usepackage{epstopdf}
\usepackage{amsmath}
\usepackage{amssymb}
\usepackage{multirow}
\usepackage{mathtools}
\usepackage[
 breaklinks=true,pdfborder={0 0 0},backref=false,colorlinks=true]
{hyperref}
\hypersetup{
 linkcolor=blue,citecolor=blue,urlcolor=blue}
\usepackage{breakurl}
\newcommand{\beq}{\begin{equation}}
\newcommand{\eeq}{\end{equation}}
\newcommand{\bea}{\begin{eqnarray}}
\newcommand{\eea}{\end{eqnarray}}

\def\ket#1{|#1\rangle}

\newcommand{\eq}{\begin{equation}}
\newcommand{\en}{\end{equation}}
\newcommand{\ear}{\begin{eqnarray}}
\newcommand{\rae}{\end{eqnarray}}

\newcommand{\rf}[1]{(\ref{#1})}
\newcommand{\scp}{\scriptsize}

\def\ket#1{|#1\rangle}

\begin{document}
\title{Universal behavior of the Shannon mutual information in non-integrable self-dual quantum chains}

\author{F.~C.~Alcaraz}

\affiliation{ Instituto de F\'{\i}sica de S\~{a}o Carlos, Universidade de S\~{a}o Paulo, Caixa Postal 369, 13560-970, S\~{a}o Carlos, SP, Brazil}


\date{\today{}}

\begin{abstract}
An existing conjecture states that the Shannon mutual information contained in the ground state 
wavefunction of conformally invariant quantum chains, on periodic lattices, has a leading 
finite-size scaling behavior that, similarly as the  von Neumann entanglement entropy, depends 
on the value of the central charge of the underlying conformal field theory describing the 
physical properties. 
This conjecture applies  whenever the ground state wavefunction is expressed in some special 
basis (conformal basis). Its formulation comes mainly from numerical 
evidences on  exactly integrable quantum chains. In this paper the above conjecture 
was tested for several general non-integrable quantum chains. We introduce new families of 
self-dual $Z(Q)$ symmetric quantum chains ($Q=2,3,\ldots$). These quantum chains contain 
nearest neighbour as well next-nearest neighbour interactions (coupling constant $p$). In the cases $Q=2$ and $Q=3$ they are extensions of the standard quantum Ising and 3-state Potts chains, 
respectively. For $Q=4$ and $Q\geq 5$ they are extensions of the Ashkin-Teller and $Z(Q)$ 
parafermionic quantum chains. Our studies indicate that these models are interesting on their 
own. They are critical,  
conformally invariant, and share the same universality class in a  continuous 
critical line. 
Moreover, our numerical analysis for $Q=2-8$ indicate that the  Shannon mutual information exhibits 
the conjectured behaviour irrespective if the conformally invariant quantum chain is exactly  
integrable or not. 
For completeness we also calculated, for these new families of quantum chains, the two existing 
generalizations of the Shannon mutual information, which are based on the R\'enyi entropy and on 
the R\'enyi divergence.

\end{abstract}
\maketitle

\section{Introduction} 
The connection between the quantum correlations and the entanglement properties 
of quantum many body systems provided us, in recent years, a powerful tool 
to detect \cite{osterloch,*PhysRevA.66.032110,*PhysRevA.70.032333,*PhysRevLett.93.250404,*PhysRevB.84.094410}
and classify quantum phase transitions (see \cite{rev-amico,*discordia,*eisert-cramer} and references therein). 
Several measures of the entanglement 
were proposed along the years, like the von Neumann and R\'enyi entanglement 
entropies \cite{eisert-cramer,vidal-etal-prl03,calabrese-cardy-jstatmech04}, the concurrence \cite{concurrence}, the fidelity \cite{fidelity-1,*fidelity-2,*fidelity-3},  etc.  Among these measures the von Neumann and R\'enyi entanglement 
entropies are the most popular since in one dimension, where most of the critical 
chains are conformally invariant, they provide a way to calculate the central charge 
of the underlying conformal field theory (CFT), identifying the  
universality class of critical behavior. Although interesting proposals were 
presented \cite{measuring-1,PhysRevLett.109.020504,
measuring-3,measuring-4} it is quite difficult to measure these quantities in the 
laboratory, and the central charge of a critical chain has never been measured 
experimentally. 

An interesting measure that is also efficient in detecting quantum phase 
transitions is the Shannon mutual information. This quantity differently from the 
previous mentioned measures is based on the measurements of observables. It 
measures the shared information among parts of a quantum system. Consider a 
quantum chain with $L$ sites that we split into two subsystems $\cal A$ and $\cal B$,
formed by consecutive $\ell$ and $L-\ell$ sites, respectively. Suppose 
the quantum chain is in the quantum state given by the wavefunction
$\ket{\Psi_{{\cal{A}}U{\cal{B}}}}=\sum_{n,m}c_{n,m}\ket{\phi_{\cal A}^n}\otimes \ket{\phi_{\cal B}^m}$,
 where $\{\ket{\phi_{\cal A}^n}\}$ and 
  $\{\ket{\phi_{\cal B}^m}\}$ are the basis spanning the subsets $\cal A$ and $\cal B$. 
The Shannon mutual information of the subsets $\cal A$ and $\cal B$ is defined as 
\beq \label{e1}
I({\cal A},{\cal B}) = Sh({\cal A}) + Sh({\cal B}) - Sh({\cal A} U {\cal B}),
\eeq
where $Sh(\chi) = - \sum_x p_x\ln{p_x}$ is the standard Shannon entropy of the subsystem $\chi$ with probability $p_x$ of being in the configuration $x$. 
The probability 
of the configurations in the subsets $\cal A$ and $\cal B$ are given by the 
marginal probabilities 
$p_{\ket{\phi_{\scriptsize{\cal A}}^n}}=\sum_m |c_{n,m}|^2$ and 
$p_{\ket{\phi_{\scriptsize{\cal B}}^m}}=\sum_n |c_{n,m}|^2$, respectively.
It is important to notice that differently from the von Neumann entanglement 
entropy and the von Neumann mutual information, which are basis independent, the 
Shannon entropy $Sh$ and the Shannon mutual information $I(\cal{A},\cal{B})$ are 
basis dependent quantities. In \cite{alcaraz-rajabpour-prl13} it was conjectured that,
 for periodic critical 
quantum chains in their 
ground state,  the Shannon mutual information shows universal features provided  the 
ground state is expressed in some special bases, called {\it conformal basis}. 
A given 
basis of the Hilbert space of the quantum chain is related to a certain  boundary condition in the time direction of the underlying (1+1)-Euclidean CFT. 
In general these time-boundary conditions destroy the conformal invariance in the bulk.
The conformal basis are related to  the boundary conditions that do not destroy 
the conformal invariance, as happens in the case of  Dirichlet and Neumann boundary 
conditions. It was conjectured \cite{alcaraz-rajabpour-prl13} that whenever the ground state 
wavefunction is expressed in the conformal basis, the leading finite-size scaling 
behavior of the Shannon mutual information for large systems and subsystem sizes
 is given by
\beq \label{e2}
I(\ell,L-\ell) = \frac{c}{4}\ln\left(\frac{L}{\pi}\sin(\frac{\ell\pi}{L})\right) + \gamma,
\eeq
where $c$ is the central charge of the underlying CFT and $\gamma$ is a 
non-universal constant. It is interesting to note that this leading behavior is the 
same as the R\'enyi entanglement entropy with R\'enyi index $n=2$ \cite{calabrese-cardy-jsm07}.

The above conjecture was tested analytically and numerically for a large number of 
exactly integrable quantum chains \cite{alcaraz-rajabpour-prl13,alcaraz-rajabpour-prb14,alcaraz-rajabpour-prb15}, namely, a set of
 coupled harmonic oscillators 
(Klein Gordon theory), the XXZ quantum chain, the Ashkin-Teller, the spin-1 
Fateev-Zamolodchikov, the $Q$-state Potts models ($Q=2,3,4$) and the $Z(Q)$
 parafermionic models ($Q=5-8$).  Up to now, except for the chain of coupled 
harmonic oscillators, this conjecture was only tested numerically. Moreover all the 
tests for this conjecture were done for exactly integrable models. Since there is no 
general analytical results supporting this conjecture it is import to check if the 
existing numerical agreement is not just a consequence of the exact integrability 
of all the  quantum chains tested so far. 
All the agreements obtained are reasonable taking into account the
lattice sizes of the considered quantum chains. However there exist a controversy in the case
of the Ising quantum chain. A numerical analysis due to
St\'ephan \cite{stephan-mutual-ising} on this quantum chain indicates that the prefactor in \rf{e2},
instead of being the central charge ($c=0.5$ in this case) is a close number $b\approx 0.4801$.
In the conclusions of this paper we present additional discussions about this point.

 In this paper we are going to check the universality feature of the conjecture 
\rf{e2} by considering critical chains belonging to several universality classes 
of critical behavior but being {\it not exactly integrable}. 

The ground state eigenfunction can only be calculated numerically for quantum 
chains of relatively small lattice sizes. It will be then interesting to consider 
non-integrable quantum chains whose critical points are  exactly known. For this sake 
we introduce in this paper a set of generalized self-dual non-integrable quantum 
chains whose exact critical points are given by  their self-dual points. Moreover, 
each of these quantum chains seems to share the same symmetries and long-distance 
physics of an exactly integrable conformally invariant chain whose central charge $c$ 
is exactly known. The validity of the conjecture \rf{e2} will imply that the 
Shannon mutual information of these models  share the same asymptotic behavior. 

We should also mention some additional studies of the Shannon and R\'enyi 
entropies and mutual information  in quantum systems \cite{PhysRevB.88.045426,*PhysRevLett.112.057203,*PhysRevB.89.165106,*PhysRevB.90.125105,*1742-5468-2014-8-P08007,*PhysRevB.93.045136,*1742-5468-2014-8-P08007,PhysRevB.93.125139}, and also in two-dimensional spin systems \cite{PhysRevE.87.022128,*1742-5468-2011-10-P10011,*1742-5468-2012-01-P01023,*PhysRevB.87.195134,*PhysRevLett.112.127204}.
The paper is organized as follows. In the next section we introduce the several 
new  quantum chains, and show their self-dual properties. In Sec.~III we present 
our results for the  models in the universality class of the Ising model and 
3-state Potts model. In Sec.~IV  the results for the models 
in the universality class of the $Z(Q)$-parafermionic models, with $Q=4,5,6,7$ and 8 
are presented. We also consider in this section a numerical analysis for a generalization of the 
$Z(Q)$ clock models with $Q=5,6,7$ and 8. 
In Sec.~V we calculate for these new quantum chains the two existing extensions 
of the Shannon mutual information: the R\'enyi mutual information and 
the less known generalized mutual information \cite{Principe:2010:ITL:1855180,alcaraz-rajabpour-prb15}. Finally in Sec.~VI we 
present our conclusions.

\section{ The $Z(Q)$ generalized self-sual quantum chains}

We introduce  initially a special generalization of the nearest-neighbor 
 Ising quantum chain that also contains 
 next-nearest neighbor interactions. The  Hamiltonian is given by:

\bea \label{e3}
&&H^{(2)}(\lambda,p) = -\sum_i 
\left[\sigma_i^z\sigma_{i+1}^z + \lambda\sigma_i^x \right.\nonumber \\
&& \left. -p(\sigma_i^z\sigma_{i+2}^z + \lambda\sigma_i^x\sigma_{i+1}^x)\right],
\eea
where $\sigma_i^z$ and $\sigma_i^x$  are spin-$\frac{1}{2}$ Pauli matrices 
attached to the lattice sites ($i=1,2,\ldots$), and $\lambda$ and $p$ are the
coupling constants. At $p=0$ the Hamiltonian \rf{e3} reduces to the standard nearest-neighbor 
quantum Ising chain, which is exactly integrable and critical at $\lambda=1$. 

In order to show that $H^{(2)}( \lambda,p)$ is self-dual, for 
any value of $p$,  let us define the 
new operators
\beq \label{e4}
\rho_{2i}^{(e)} = \sigma_i^z\sigma_{i+1}^z \mbox{  and  } \rho_{2i-1}^{(o)} = 
\sigma_i^x, \quad i=1,2,\ldots,
\eeq
that obey the following commuting and anti-commuting relations
\bea \label{e5}
&&\left({\rho_i^{(e)}}\right)^2=
\left({\rho_i^{(o)}}\right)^2=1, \quad 
[\rho_i^{(o)},\rho_j^{(o)}]=
[\rho_i^{(e)},\rho_j^{(e)}]=0,\nonumber \\
&&[\rho_i^{(o)},\rho_j^{(e)}]=0, \mbox{  unless  } |i-j|=1, \nonumber \\
&&\{\rho_i^{(e)},\rho_j^{(o)}\}=0, \mbox{ if } |i-j|=1.
\eea
In terms of these new operators the Hamiltonian \rf{e3} is given by
\bea \label{e6}
&&H^{(2)}(\lambda,p) = -\sum_i\left[\rho_{2i}^{(e)}+\lambda\rho_{2i-1}^{(o)} \nonumber \right.\\
+&&
\left. p(\rho_{2i}^{(e)}\rho_{2i+2}^{(e)}+ \lambda\rho_{2i-1}^{(o)}\rho_{2i+1}^{(o)})\right].
\eea

We now make a transformation by defining the new operators:
\beq \label{e7}
\tilde{\rho}_{2i}^{(e)}=\rho_{2i+1}^{(o)}, \quad 
\tilde{\rho}_{2i-1}^{(o)}=\rho_{2i}^{(e)}.
\eeq
It is simple to see that these new operators obey the same commutation relations as 
the old ones, given in \rf{e5}. In terms of these new operators the Hamiltonian \rf{e3} is now given by
\bea \label{e8}
&&H^{(2)}(\lambda,p) = -\lambda \sum_i[
\tilde{\rho}_{2i}^{(e)} + \frac{1}{\lambda} \tilde{\rho}_{2i-1}^{(o)} \nonumber \\
&&+p (\tilde{\rho}_{2i}^{(e)}\tilde{\rho}_{2i+2}^{(e)}  + \frac{1}{\lambda} \tilde{\rho}_{2i-1}^{(o)}\tilde{\rho}_{2i+1}^{(o)})]. 
\eea
Consequently, apart from a boundary term \footnote{This transformation for finite lattices will produce constraints among the operators $\{\rho_i^{(o)},\rho_i^{(e)}\}$ 
and the exact relation for finite chains only relate sectors of the 
associated Hilbert space.} \cite{abb} that could be neglected as the lattice size 
increases, the model is self-dual:
\beq \label{e9}
H^{(2)}(\lambda,p) = \lambda H^{(2)}(\frac{1}{\lambda},p).
\eeq
Implying that the low-lying eigenlevels in the eigenspectrum of both 
sides of \rf{e9}  become identical as 
the lattice size increases.  Since we have no reason to expect more than a single 
$Z(2)$ critical point for a fixed value of $p$, this model should be critical at $\lambda=1$ and {\it at least} for $p\leq p_c$ (with $p_c$ finite) the model should share 
the same universality class as the standard quantum Ising chain $H^{(2)}(1,0)$. 
Actually for $p \to \infty$ the model is $Z(2)\otimes Z(2)$ symmetric due to the 
commutations of $H^{(2)}(\lambda,p\to \infty)$ with the nonlocal $Z(2)$ 
operators ${\cal{P}}^{(e)} = \prod_{i}  \sigma_{2i}^x$ and 
 ${\cal{P}}^{(o)} = \prod_{i}  \sigma_{2i-1}^x$, and therefore is  not  in the Ising 
universality class.

Similarly as we did for the Ising quantum chain we now introduce the self-dual 
generalized next-nearest neighbor $Z(Q)$ models ($Q=2,3,\ldots$). They describe the 
dynamics of the $Q\times Q$ matrices $\{S_i\}$, $\{R_i\}$, attached on the lattice 
sites $i=1,2,\ldots$, and obey the algebraic relations
\bea \label{e10}
&&S_i^Q=R_i^Q=1,\quad [S_i,S_j]=[R_i,R_j]=0, \nonumber \\
&&[S_i,R_j]=0 \mbox{ if } i\neq j\mbox{ and } S_iR_i=e^{i\frac{2\pi}{Q}}R_iS_i.
\eea
The Hamiltonian we introduce  is given by
\bea \label{e11}
&&H^{(Q)}(\lambda,\{\alpha\}) = -\sum_i \left[
\sum_{n=1}^Q \alpha_n(S_i^nS_{i+1}^{Q-n} + \lambda R_i^n) \right. \nonumber \\
&& \left. + p \sum_{n=1}^Q \alpha_n(S_i^nS_{i+2}^{Q-n} + \lambda R_i^nR_{i+1}^n)\right],
\eea
where $\lambda$ and $\{\alpha_n\}$ ($n=1,\ldots,Q$) are coupling constants. We chose 
real coupling constants and  $\alpha_n=\alpha_{Q-n}$ to ensure the hermiticity 
of the Hamiltonian. This Hamiltonian reduces to \rf{e3} for $Q=2$. 

We now consider the $Z(Q)$ operators:
\beq \label{e12}
\rho_{2i}^{(e)}=S_iS_{i+1}^{Q-1} \mbox{ and } \rho_{2i-1}^{(o)}=R_i, \quad i=1,2,\ldots,
\eeq
that obey the following algebraic relations
\bea \label{e13}
&&\left(\rho_i^{(e)}\right)^Q=
\left(\rho_i^{(o)}\right)^Q=1, 
[\rho_i^{(e)},\rho_j^{(e)}]=
[\rho_i^{(o)},\rho_j^{(o)}]=0, \nonumber \\
&&[\rho_i^{(e)},\rho_j^{(o)}]=0 \mbox{ unless } |i-j|=1,      \nonumber \\
&&\rho_i^{(e)}\rho_{i\pm 1}^{(o)}=e^{\mp i\frac{2\pi}{Q}}  
\rho_{i\pm 1}^{(o)}\rho_{i}^{(e)}.
\eea 
In terms of these operators we have
\bea \label{e14}
&&H^{Q)} (\lambda,\{\alpha\}) = -\sum_i\left\{ \sum_{n=1}^{Q-1}\alpha_n \left[
(\rho_{2i}^{(e)})^n + \lambda (\rho_{2i-1}^{(o)})^n \right] \right. \nonumber \\ 
&& \left. +p\sum_{n=1}^{Q-1} \alpha_n \left[ 
(\rho_{2i}^{(e)} \rho_{2i+2}^{(e)})^n 
+\lambda ( \rho_{2i-1}^{(o)}\rho_{2i+1}^{(o)} )^n\right] \right\}.
\eea
We now perform the same canonical transformation $\rho \to \tilde{\rho}$, given by 
\rf{e7}. It is simple to verify that the 
transformation is canonical since the commutation's relations of the new 
operators are the same as the old ones. 
The Hamiltonian is now given by:
\bea \label{e15}
&& H^{(Q)}(\lambda,\{\alpha\}) = - \lambda \left\{ \sum_{n=1}^{Q-1}  
\alpha_n \left[(\tilde{\rho}_{2i}^{(e)})^n  
+ \frac{1}{\lambda}( \tilde{\rho}_{2i-1}^{(o)} )^n \right]  \right. \nonumber \\
&& \left. +p\sum_{n=1}^{Q-1}\alpha_n \left[  
(\tilde{\rho}_{2i}^{(e)}\tilde{\rho}_{2i+2}^{(e)})^n +  
\frac{1}{\lambda} ( \tilde{\rho}_{2i-1}^{(o)}\tilde{\rho}_{2i+1}^{(o)} )^n \right] \right\}. 
\eea  
Comparing \rf{e14} and \rf{e15}, we obtain, apart
 from a boundary term [18] 
\beq \label{e16}
H^{(Q)} (\lambda,\{\alpha\}) = \lambda 
H^{(Q)} (\frac{1}{\lambda},\{\alpha\}).
\eeq
The particular choice $\alpha_n = \frac{1}{\sin(\frac{\pi n}{Q})}$, 
$n=1,2,\ldots,Q-1$ give us an interesting family of quantum chains that we are 
going to study in the next sections. At their  self-dual point ($\lambda=1$) these 
Hamiltonians are given by:
\bea \label{e17}
&&H^{(Q)}(p) = -\sum_i\left\{\sum_{n=1}^{Q-1} \frac{1}{\sin(\frac{\pi n}{Q})}\left[
S_i^nS_{i+1}^{Q-n} + R_i^n \right. \right. \nonumber \\
&& \left. \left.+ p(S_i^nS_{i+2}^{Q-n} + R_i^n R_{i+1}^n)\right] \right\} . 
\eea
These Hamiltonians at $p=0$ are critical, conformal invariant and 
exactly integrable. They correspond for $Q=2,3$ to the 2-state and 3-state Potts 
models, for $Q=4$ it is the Ashkin-Teller model with a special value of 
its anisotropy, and for $Q>4$ they correspond to the $Z(Q)$ parafermionic models 
\cite{fateev-paraf,*alcaraz-lima-1986,*alcaraz-fssb}. For $p\neq 0$ the models lose their exact integrability but we do expect 
that, at least for small values of the parameter $p$, they stay critical and in the 
same universality class of the related $p=0$ exactly integrable quantum chain. For 
large values of $p$ this may not be true since, as happened in the 
Ising case, for $p \to \infty$  the symmetry increases from a single $Z(Q)$ to 
 a $Z(Q)\times Z(Q)$.

\section{Results for the extended Ising and 3-state Potts quantum chains}

We present in this section our numerical results for the generalized self-dual 
Ising and 3-state Potts quantum chains whose Hamiltonians $H^{(Q)}(p)$ are 
given by \rf{e17} with the values $Q=2$ and $Q=3$, respectively. At $p=0$ these 
models are exactly integrable and conformally invariant, being ruled by a CFT 
 with central charge $c=1/2$ and $c=4/5$, respectively. Our aim is to compute 
the Shannon mutual information for the values of the parameter ($p\neq 0$) where the models are still critical but not exactly integrable. Since we are testing a 
conjecture we should initially confirm the expectation  that the models, for small values of 
the parameter $p$ are still critical and in the same universality class as 
the $p=0$ exactly integrable quantum chain. 

A first test of the critical universality for the quantum chains can be done 
by comparing  their central charge  $c$ calculated directly from the finite-size 
behavior of the ground state energy and low-lying energy gaps.  The 
ground state energy $E_0(L)$ of a conformally invariant quantum chain with periodic 
boundary should have the asymptotic behavior \cite{cardy-anomaly,*affleck-anomaly}:
\beq \label{e18}
\frac{E_0}{L} = e_{\infty} - v_s \frac{\pi c}{6L^2}  + o(L^{-2}),
\eeq
where $e_{\infty}$ is the energy per site in the bulk limit and $v_s$ is 
the sound velocity. 
	The sound velocity can be extracted from the leading finite-size 
behavior of the first energy gap related to a given primary operator of the 
underlying CFT \cite{cardy-fss-1984,*cardy-operator-content-1}. For example the lowest energies $E_1(p)$ in the 
eigensector with $Z(Q)$ charge $q=1$ and momentum $P=0,\frac{2\pi}{L},
\frac{4\pi}{L},\ldots$ are associated to the $Z(Q)$-magnetic operators  of 
these models. We have then the estimate $v_s(L)$ for the sound velocity 
 \cite{gehlen1}
\beq \label{e19}
v_s(L)= \frac{L[E_1(\frac{2\pi}{L}) - E_1(0)]}{2\pi} + o(L^{-1}), \nonumber 
\eeq
that together with \rf{e18} give us an estimate for the central charge 
of the quantum chain:
\beq \label{e20}
c_{\mbox{\scriptsize{est}}}(L) = 
-\frac{\frac{E_0(L)}{L}-\frac{E_0(L-1)}{L-1}}
{\frac{1}{L^2}-\frac{1}{(L-1)^2}} \frac{12}{L(E_1(\frac{2\pi}{L})-
E_1(0))}  + o(L^{-1}).
\eeq

In Fig.~1 and ~2 we illustrate our results for the estimate 
$c_{\scp \mbox{est}}(L)$ in the extended self-dual Ising and 3-state Potts 
models, respectively. We consider the models with the parameter 
$p=0,0.5,1$ and 1.5, and lattice sizes up to $L_{\scp \mbox{max}}=30$ for the Ising case and 
$L_{\scp \mbox{max}}=19$ for the 3-state Potts case. We also show in the 
figures the estimated results $c_{\scp \mbox{est}}(L\to \infty)$ for 
the central charge $c$. They were obtained by considering a simple 
quadratic fit of $c_{\scp \mbox{est}}(L)$ for $30\leq L\leq 20$ in the 
Ising case and $19\leq L\leq11$ in the 3-state Potts case. The numerical 
results in these figures indicate that for the parameters $p\lesssim 1.5$ the 
extended models stay in the same universality class of the related $p=0$ 
exactly integrable model, i. e., $c=1/2$ ad $c=8/10$ for the Ising and 
3-sate Potts models, respectively.

\begin{figure} [htb] \label{fig1}
\centering
\includegraphics[width=0.35\textwidth]{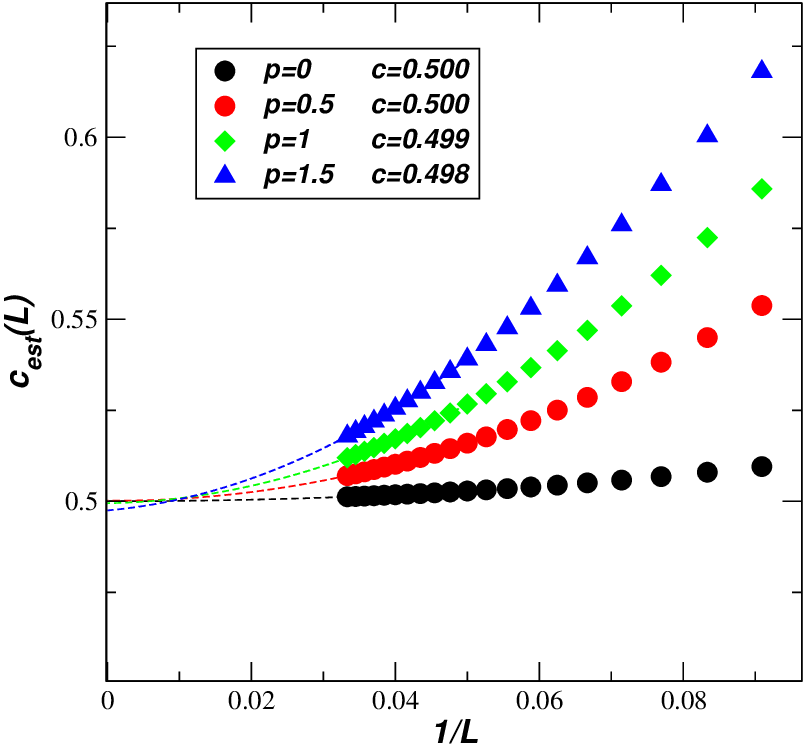}
\caption 
{The estimate $c_{\scp \mbox{est}}(L)$ given by \rf{e20} as a function of 
$1/L$ for the extended self-dual Ising model given by the Hamiltonian \rf{e3}, and for the values of the parameter $p=0,0.5,1$ and 1.5. The estimated values 
$c_{\scp \mbox{est}}(L\to\infty)=c$, shown in the figure, were obtained 
from a quadratic fit by considering the lattice sizes $20\leq L \leq 30$.}
\end{figure}
\begin{figure} [htb] \label{fig2}
\centering
\includegraphics[width=0.35\textwidth]{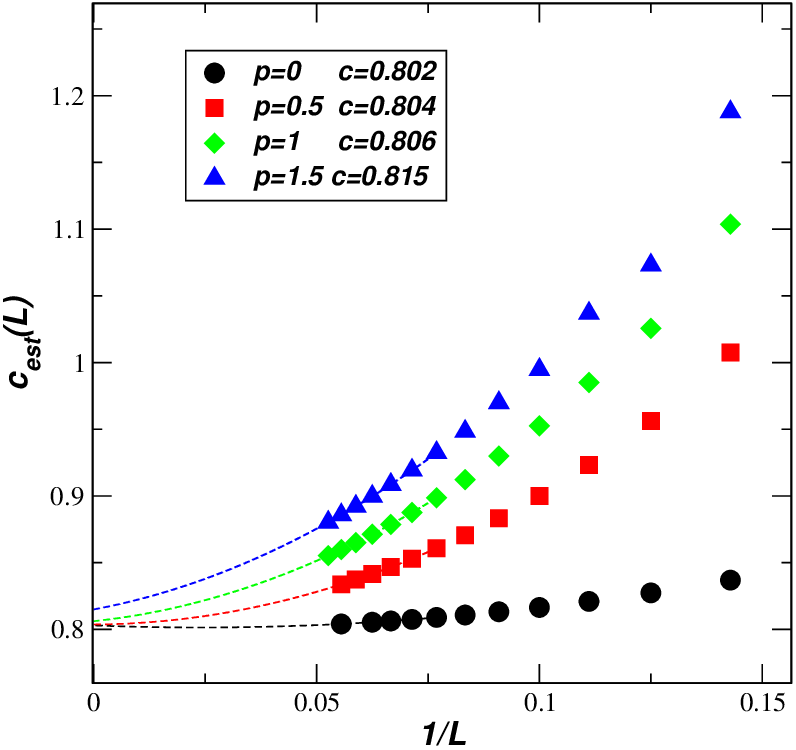}
\caption
{The estimate $c_{\scp \mbox{est}}(L)$ given by \rf{e20}, as a function of 
$1/L$, for the extended 3-state Potts quantum chain  by the Hamiltonian 
\rf{e17} and for the values of the parameter $p=0,0.5,1$ and 1.5. The estimated values 
$c_{\scp \mbox{est}}(L\to\infty)=c$, shown in the figure, were obtained 
from a quadratic fit by considering the lattice sizes $14\leq L \leq 19$.}
\end{figure}

A second test can be done by calculating  the von Neumann entanglement 
entropy $S_{vN}(\ell,L)$ of subsystems with sizes $\ell$ and $(L-\ell)$ in the quantum 
chains. Its finite-size scaling behavior, for a periodic chain, is giving by 
\cite{holzhey-larsen-wilczek,korepin-entropy-1,calabrese-cardy-jpa-09}
\beq \label{e21}
S_{vN}(\ell,L-\ell) = \frac{c}{3}\ln(\frac{L}{\pi}\sin(\frac{\ell\pi }{L})) + k,
\eeq
where $k$ is a constant. 
In order to calculate $S_{vN}(\ell,L)$, from a given ground state wave function,  
we should fully diagonalize the reduced density matrix of the subsystems 
(dimension $Q^{\ell} \times Q^{\ell}$). This brings an extra numerical 
limitation since we can only handle the complete  diagonalization of matrices 
with dimensions smaller than $\sim$6000.
  We are then restricted for the $Q=2$ ($Q=3$) model with sublattices 
sizes $\ell \leq 12$  ($\ell \leq 7$). 

In Fig.~3 (Fig.~4) we show, for 
several values of $p$, $S_{vN}(\ell,L)$ as a function of 
$\sin(\frac{L}{\pi}\sin(\frac{\pi}{L}))/3$ for the $Q=2$  ($Q=3$) extended quantum chains with $L=24$ ($L=14$) sites. It is  also shown in these figures the estimated values of the central 
charge obtained from a linear fit.  These results clearly indicate that these quantum chains are 
indeed critical, and share the same universality class of critical behavior 
as the exactly integrable quantum chain  $p=0$, whose central charge is $c=0.5$.
\begin{figure} [htb] \label{fig3}
\centering
\includegraphics[width=0.35\textwidth]{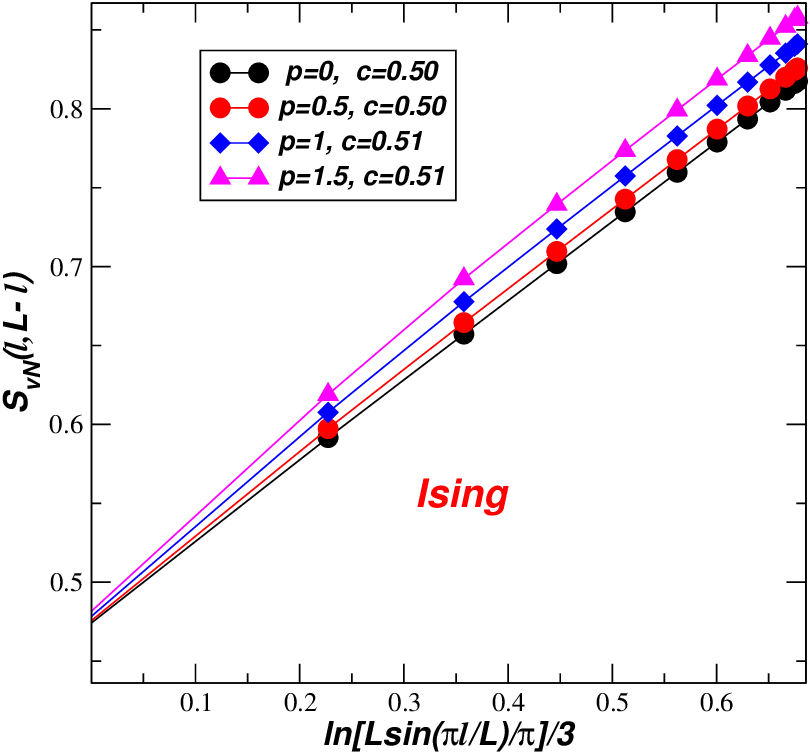}
\caption
{The von Neumann entropy for the extended self-dual Ising model \rf{e3} with 
 $L=24$ sites and the parameter values $p=0,0.5,1$ and 1.5. The estimated values for the central 
charge are shown. They were obtained from a linear fit (see \rf{e21}), considering    
the 
sublattice sizes $\ell=5-12$.} 
\end{figure}
\begin{figure} [htb] \label{fig4}
\centering
\includegraphics[width=0.35\textwidth]{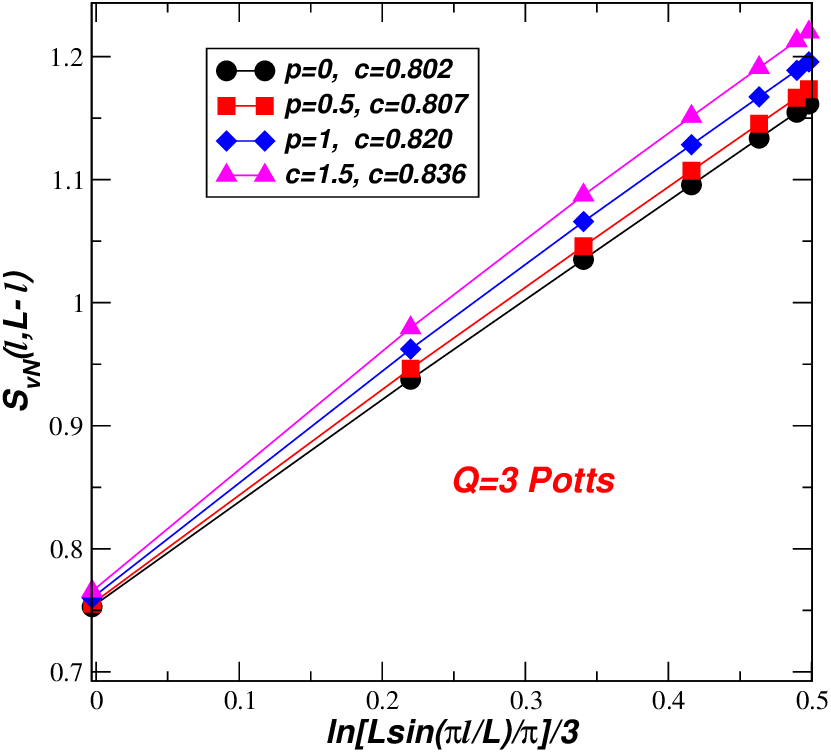}
\caption
{The von Neumann entropy for the extended 3-state Potts model \rf{e17} with 
$L=14$ sites and the  values of the parameter $p=0,0.5,1$ and 1.5. The estimated values for the central 
charge are shown. They were obtained from a linear fit (see \rf{e20}), considering  the 
sublattice sizes $\ell=4-7$.} 
\end{figure}

Once we have convinced ourselves about the universal behavior of these 
non-integrable quantum chains for $0\leq p \lesssim 1.5$, we can now test the 
universal behavior \rf{e2} claimed for the Shannon mutual information 
$I(\ell,L-\ell)$ of periodic quantum chains in their ground states. 

The Shannon mutual information depends on the particular basis we chose to 
express the ground state weave function. The previous results \cite{alcaraz-rajabpour-prl13,alcaraz-rajabpour-prb14}, based on exactly 
integrable quantum chains, indicate that  two good basis, where the universal
 behavior are shown, are the basis where either the "kinetic interactions" or 
the "static interactions" are diagonal. In the set of models we are testing 
these  basis are the ones where  the operators $\{S_i\}$  
 or $\{R_i\}$ are diagonal.

In Fig.~5 and Fig.~6 the Shannon mutual information are 
shown for the extended Ising chain \rf{e3} with $L=30$ sites and for  values of 
the parameter $p=0,0.5,1$ and $1.5$. The results of Fig~5 (Fig.~6) are 
obtained from the  
ground state wavefunction given in the $\{\sigma^z\}$-basis 
($\{\sigma^x\}$-basis). 
We clearly see in these figures a linear behavior indicating 
$\ln(L\sin(\pi\ell/L))$ as the finite-size scaling function. 
The  estimated values of the central charge 
$c =0.48-0.50$,  are also close to the expected value $c=1/2$.
These 
estimates were obtained from a linear fit by considering all  
the sublattice sizes.
\begin{figure} [htb] \label{fig5}
\centering
\includegraphics[width=0.35\textwidth]{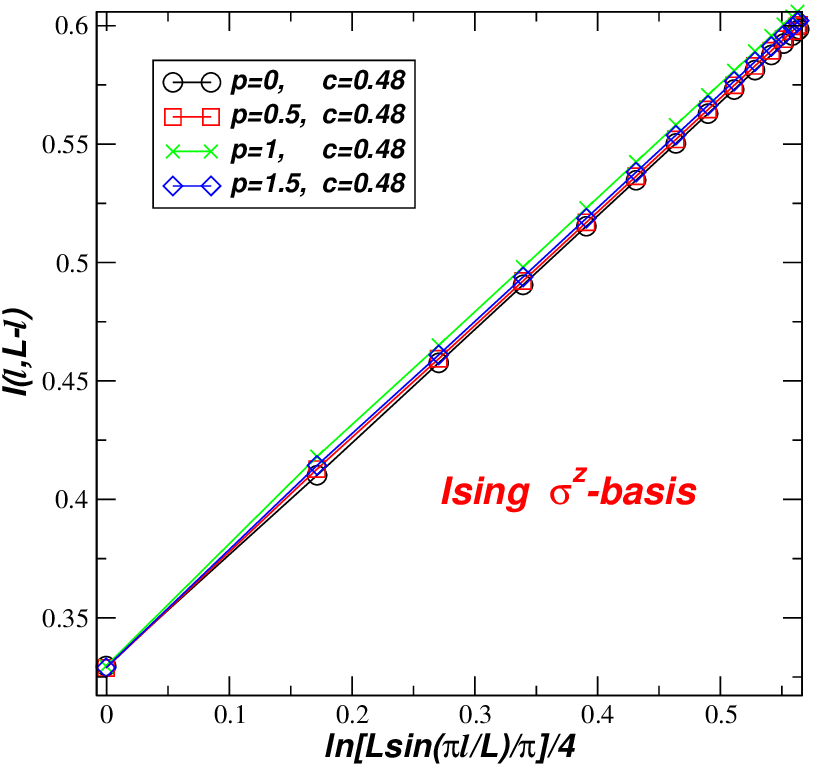}
\caption
{The Shannon mutual information $I(\ell,L-\ell)$, as a function of 
$\ln[L\sin(\pi\ell/L)/\pi]/4$, for the extended selfdual Ising quantum chain 
\rf{e3}, with the values of the parameter $p=0, 0.5,1$ and 1.5. 
The results are obtained for the ground state wavefunction of the $L=30$ sites 
quantum chain expressed in the basis where  $\{\sigma_i^z\}$ are diagonal. 
The estimated 
results, based on the conjecture \rf{e2} are also shown. They were obtained 
from a linear fit by considering all the sublattices sizes.}
\end{figure}
\begin{figure} [htb] \label{fig6}
\centering
\includegraphics[width=0.35\textwidth]{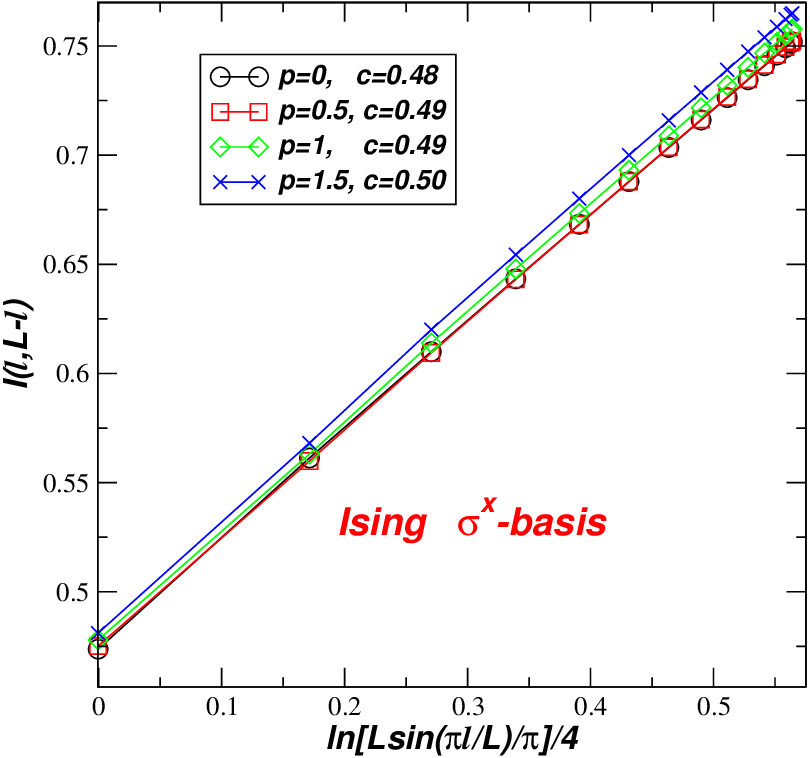}
\caption{ Same as in Fig.~5 but with the ground  state wavefunction expressed 
in the basis where $\{\sigma_i^x\}$ are diagonal.}
\end{figure}

In Fig.~7 and Fig.~8 we show the Shannon mutual information 
for the extended $Z(3)$ models with the values of the parameter $p=0,0.5,1$ and $1.5$. 
In Fig.~7 (Fig.~8) the quantum chain has $L=18$ 
($L=19$) sites and is in the basis where the matrices $\{S_i\}$ ($\{R_i\}$) are diagonal, respectively. 
 The 
linear fit obtained by using all the sublattice sizes predicts the value 
for the central charge $c\approx 0.77-0.79$. These values are   close to the predicted 
value $c=8/10$, indicating the validity of the conjecture \rf{e2} even for 
non-integrable quantum chains. 
It is interesting to notice that differently from the calculation 
 of $S_{vN}(\ell,L)$, it is not necessary to full diagonalize reduced 
matrices and we could calculate $I(\ell,L-\ell)$ for larger lattice sizes, 
namely $L=30$ and $L=19$ for the extended Ising and 3-state Potts chains, 
respectively. 
\begin{figure} [htb] \label{fig7}
\centering
\includegraphics[width=0.35\textwidth]{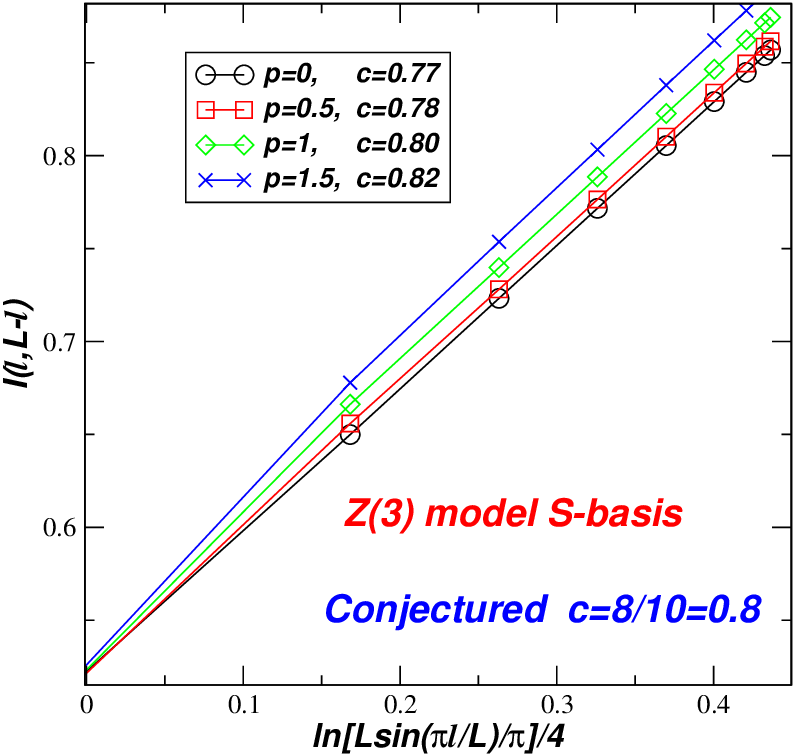}
\caption
{The Shannon mutual information $I(\ell,L-\ell)$, as a function of 
$\ln[L\sin(\pi\ell/L)/\pi]/4$, for the extended $Q=3$ selfdual Potts quantum chain 
\rf{e17}, with the values of the parameter $p=0, 0.5,1$ and 1.5. 
The results are obtained for the ground state wavefunction of the $L=18$ sites 
quantum chain, expressed in the basis where  $\{S_i^z\}$ are diagonal. 
The estimated 
results, based on the conjecture \rf{e2} are also shown. They were obtained 
from a linear fit by considering all the sublattice sizes.}
\end{figure}
\begin{figure} [htb] \label{fig8}
\centering
\includegraphics[width=0.35\textwidth]{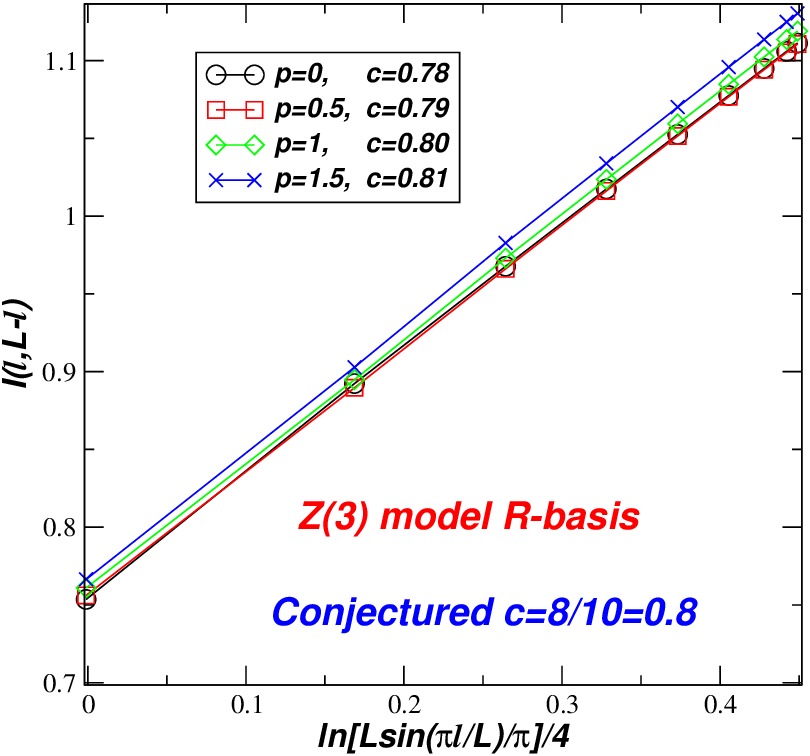}
\caption{ Same as Fig.~7 but for lattice size $L=19$ and the results are 
obtained from the ground state wavefunction expressed in the $\{R_i\}$ basis.}
\end{figure}

\section{Results for the extended Z(Q)-parafermionic quantum chains} 

We consider in this section the numerical tests of the conjecture \rf{e2} for the extended non-integrable 
$Z(Q)$-parafermionic models \rf{e17}. The cases where the parameter 
$p=0$ reduces to the known exactly integrable $Z(Q)$-parafermionic 
quantum chains \cite{fateev-paraf,*alcaraz-lima-1986,*alcaraz-fssb}, which are critical and conformally invariant with 
conformal central charges:
\beq \label{e22}
c= \frac{2(Q-1)}{Q+2}, \quad Q=2,3,\ldots \, .
\eeq
The cases $Q=2$ and $Q=3$ are the Ising and 3-state Potts models considered 
in the last section. The quantum chain with $Q=4$ 
 corresponds to a particular anisotropy of the $c=1$ critical line of the 
quantum  Ashkin-Teller chain. 
 The 
cases $Q>4$ are the Z(Q)-parafermionic quantum chains with central charge 
$c>1$. Actually these last models are multicritical points and are expected 
to be endpoints \cite{alcaraz-fssb,alcaraz-fssc} of critical lines belonging to a massless phases with central 
charge $c=1$ and belonging to the Berezinskii-Kosterlitz-Thouless 
universality class \cite{alcaraz-fssc,clock-6states}. 

The Shannon mutual information for the extended $Q=4$ quantum chain with 
the values of $p=0,0.5,1$ and $1.5$ are shown in Fig.~9. The calculations 
were done by expressing the ground state wavefunction either in the $S$-basis 
($L=14$) or in the $R$-basis ($L=13$). The linear fit, using all the 
 sublattice sizes, give the estimated values of the central charge shown in 
the figure $c\approx 0.97- 1.03$,  which within the numerical accuracy 
corroborates the conjecture \rf{e2}. 
\begin{figure} [htb] \label{fig9}
\centering
\includegraphics[width=0.35\textwidth]{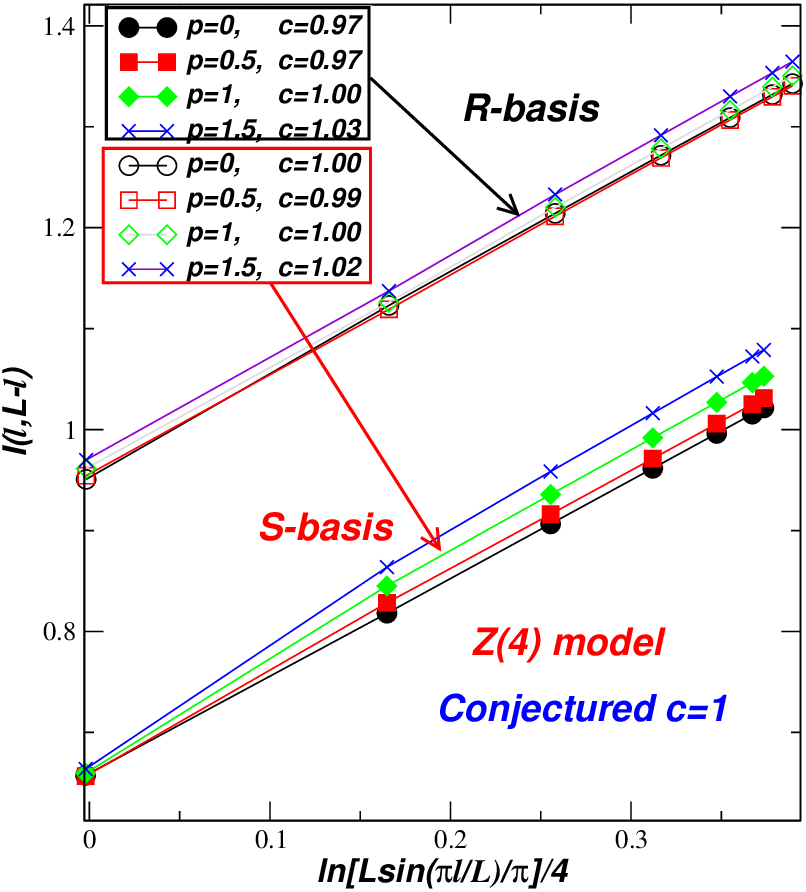}
\caption{ 
 The Shannon mutual information $I(\ell,L-\ell)$, as a function of 
$\ln[L\sin(\pi\ell /L)/\pi]/4$ for the extended $Q=4$ self-dual quantum 
chain \rf{e17}, with the values of the parameters $p=0,0.5,1$ and 1.5. The 
results were obtained for the lattice size $L=15$ and $L=14$, when the 
ground state wavefunction spanned  in the basis where $\{S_i\}$ and $\{R_i\}$ are diagonal, 
respectively. The estimated values shown in the figure were obtained from a 
linear fit by considering all the sublattice sizes. }
\end{figure}

Let us now consider the extended models with $Q>4$. Since the $p=0$ models 
are multicritical it is not clear if the non-integrable quantum chains, 
although critical, will stay in the same universality class as the integrable model $p=0$. 
{\it Surprisingly} this seems to be  the case. In Figs.~10,~11,~12 and ~13  we show for some 
values of $p$ the Shannon mutual information for the quantum chains with 
$Q=5,6,7$ and 8, respectively. The calculation were done for the ground state 
wavefunction expressed in the basis where either $\{S_i\}$ or $\{R_i\}$ are diagonal.  
The lattice sizes used are given in the figure captions. The estimated values for the 
central charge are givem in the figure and were obtained from a linear 
fit, where all the sublattice sizes are considered.  They are close to the 
predicted values: $c=8/7=1.14285...$ ($Q=5$), $c=5/4=1.25$ ($Q=6$), 
$c=4/3=1.333...$ ($Q=7$) and $c=7/5=1.4$ ($Q=8$).  Taking into account the lattice sizes we could calculate, 
these results indicate that the models are still in the same universality class of the 
multicritical point ($p=0$), at least for the values of parameters   $0<p \lesssim 1$. These 
results tests the universal character of the conjecture \rf{e2}, 
corroborating its validity for non-integrable critical quantum chains. 
\begin{figure} [htb] \label{fig10}
\centering
\includegraphics[width=0.35\textwidth]{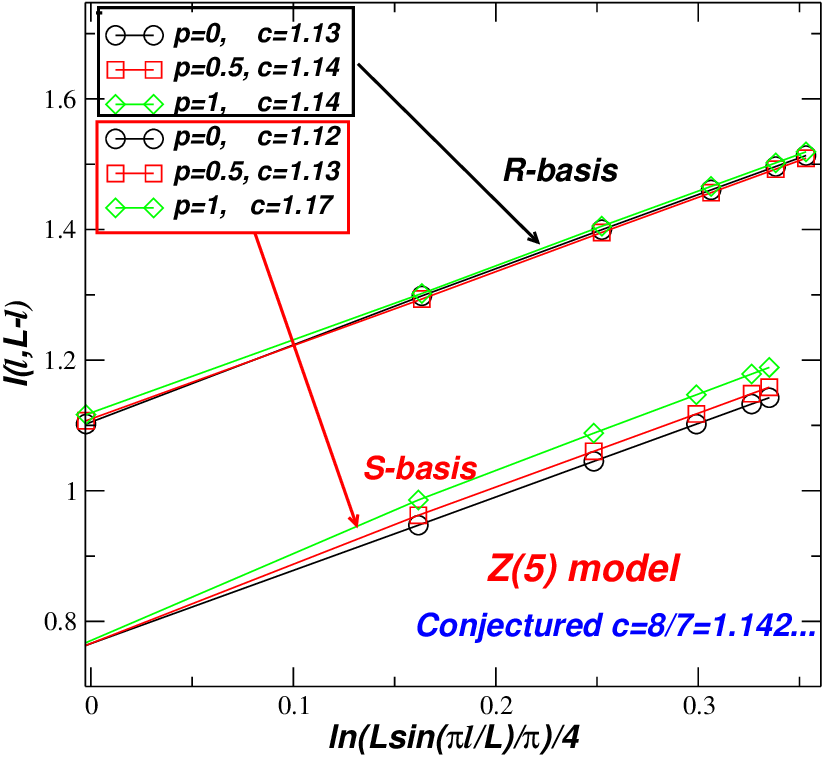}
 \caption{The Shannon mutual information $I(\ell,L-\ell)$, as a function of 
$\ln[L\sin(\pi\ell /L)/\pi]/4$ for the extended $Q=5$ self-dual quantum 
chains \rf{e17}, with the values of the parameters $p=0,0.5$ and 1. The 
results were obtained for the lattice sizes $L=12$ and $L=13$, when the 
ground state wavefunction are in the basis where $\{S_i\}$ and $\{R_i\}$ are diagonal, 
respectively. The estimated values shown in the figure were obtained from a 
linear fit by considering all the sublattice sizes.}
\end{figure}
\begin{figure} [htb] \label{fig11}
\centering
\includegraphics[width=0.35\textwidth]{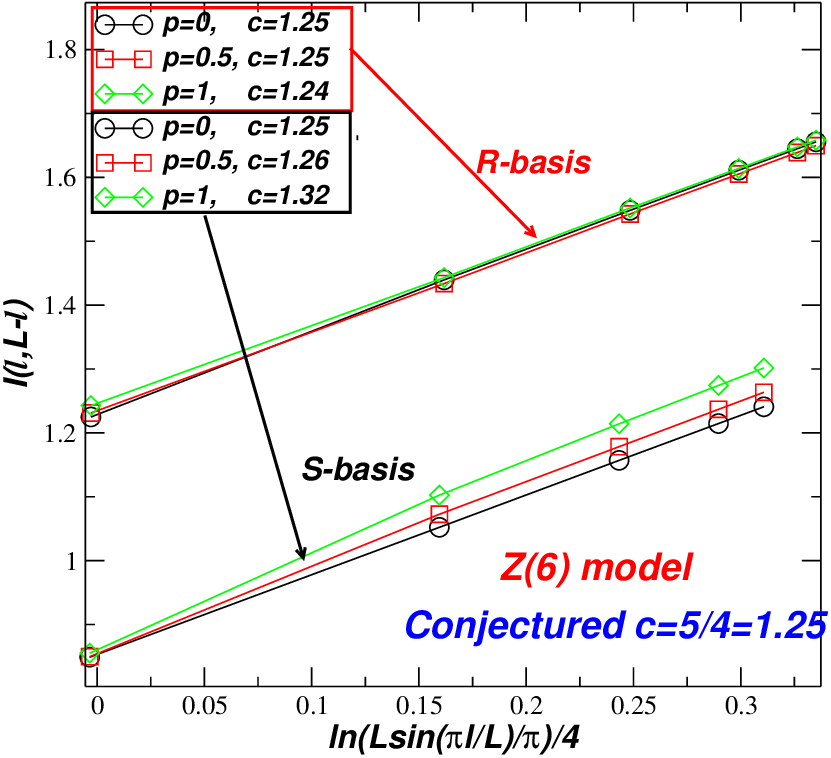}
\caption{ Same as Fig.~10 for the extended $Z(6)$ sef dual quantum 
chain \rf{e17}. The lattice sizes are $L=12$ and $L=13$ for the basis where 
$\{S_i\}$ and $\{R_i\}$ are diagonal, respectively.}  
\end{figure}
\begin{figure} [htb] \label{fig12}
\centering
\includegraphics[width=0.35\textwidth]{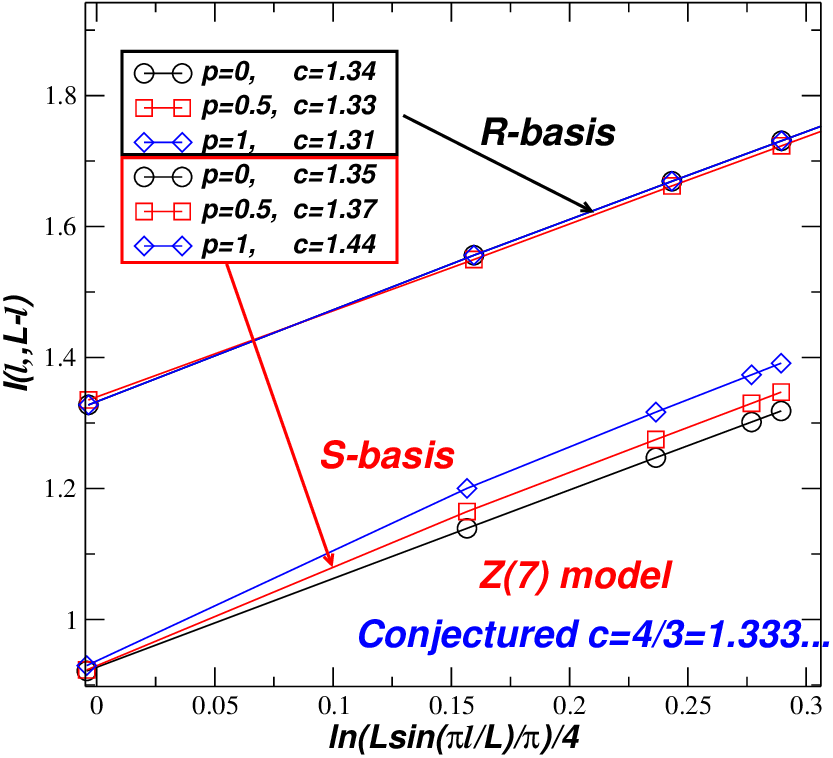}
\caption{ Same as Fig.~10 for the extended $Z(7)$  self-dual quantum 
chain \rf{e17}. The lattice sizes are $L=11$ and $L=12$ for the basis where 
$\{S_i\}$ and $\{R_i\}$ are diagonal, respectively.}  
\end{figure}
\begin{figure} [htb] \label{fig13}
\centering
\includegraphics[width=0.35\textwidth]{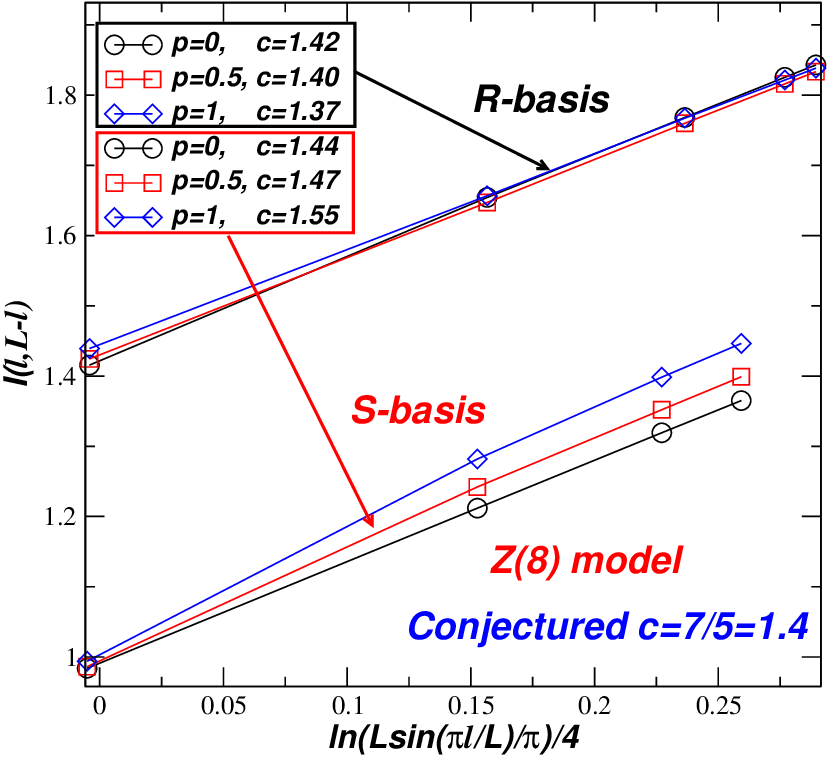}
\caption{ Same as Fig.~10 for the extended $Q=8$ self-dual quantum 
chain \rf{e17}. The lattice sizes are $L=10$ and $L=11$ for the basis where 
$\{S_i\}$ and $\{R_i\}$ are diagonal, respectively.}  
\end{figure}

Before closing this section let us do an additional test for the conjecture \rf{e2}. 
For $Q\geq5$ the $Z(Q)$ family of clock quantum chains (which is  related to the time-continuum limit of the 2-d classical clock models \cite{jose-kadanoff})  is known to have, besides a disordered and ordered phases, 
an intermediate  massless phase 
belonging to the Berezinskii-Kosterlitz Thouless universality and are expected 
to be ruled by a CFT with  central charge $c=1$ \cite{alcaraz-fssc,clock-6states}. These models,
 although not exactly integrable, are
self-dual. Their self-dual points belong to the intermediate $c=1$ CFT. Exploring 
the general results of Sec.~2, similarly as we did for the Z(Q) parafermionic models, 
we can extend the standard clock models by choosing in \rf{e11} 
$\alpha_n=\delta_{n,1} + \delta_{n,Q-1}$ for ($n=1,\ldots,Q-1$). At its self-dual 
point the extended clock models are given by
\bea \label{e22a}
&&H_{\scp\mbox{clock}}(p) = -\sum_i\left[ 
S_iS_{i+1}^+ + S_i^+S_{i+1} + R_i + R_i^+  +  \right. \nonumber \\
&& \left. p(S_iS_{i+2}^+ + S_i^+S_{i+2} + R_iR_{i+1}+R_i^+R_{i+1}^+)\right],
\eea
where, as before, $S_i$ and $R_i$ are the $Z(Q)$ matrices with algebraic relations 
given by \rf{e10}. At $p=0$ these Hamiltonians reduce to the standard $Z(Q)$ clock 
quantum chains. Our numerical results indicate that for arbitrary values of $0\leq p \leq 1$ 
the models share the same $c=1$ CFT. In Fig.~14 we show our tests for the Shannon mutual 
information $I(\ell,L-\ell)$ for the $Z(Q)$ clock model with $Q=5,6,7$ and 8. We only present 
the results in the case where the ground state wavefunction is expressed in the $\{R_i\}$ basis. In this figure, 
for each value of $Q$ the data are for the values of the parameter $p=0,0.5$ and 1. We clearly 
see the linear dependence with $\ln[L\sin(\pi\ell/L)]/4$. The linear fit, by considering 
all the values of $p$, and sublattice sizes for a given $Z(Q)$ model, give us estimates of 
the central charge in the range $c=1.03-1.04$, that are close to the expected value $c=1$, 
indicating the validity of the conjecture \rf{e2}. 
\begin{figure} [htb] \label{fig14}
\centering
\includegraphics[width=0.35\textwidth]{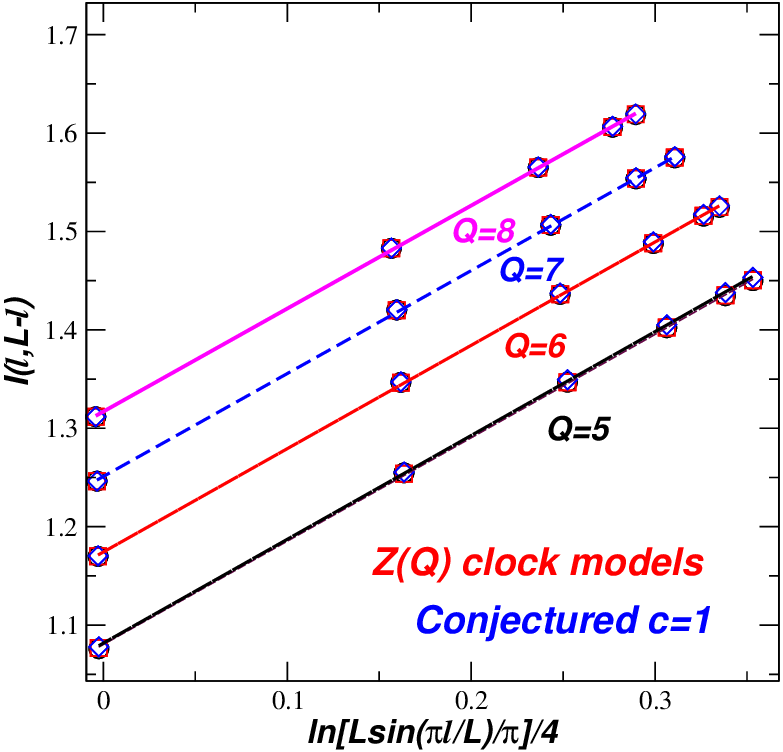}
\caption{ 
The Shannon mutual information for the extended $Z(Q)$ clock models defined in \rf{e22a}, for the 
values of $Q=5,6,7$ and 8, and lattice sizes $L=13,12,11$ and 10, respectively. For each 
$Z(Q)$ model the results are for the values of the parameter $p=0,0.5$ and 1. The calculations 
were done for the ground state spanned in the $\{R_i\}$ basis. The  lines are the 
linear fit considering all the points for a given $Z(Q)$ model. }
\end{figure}

\section{ Generalized mutual informations}

A crucial step in deriving most of the analytical results (e.g. \cite{calabrese-cardy-jpa-09,PhysRevLett.106.201601,*1742-5468-2012-01-P01016}) for the 
von Neumann entanglement entropy come from two facts. The Shannon entropy 
is obtained from the $n\to 1$ limit of the $n$-R\'enyi entanglement entropy, 
and at this limit the replica trick, used for the conformal transformations, 
is regular. There exists two generalizations of the Shannon mutual 
information considered in the early sections. These extensions are based 
either on the R\'enyi entropy or on the R\'enyi divergence \cite{Principe:2010:ITL:1855180}. Previous 
 numerical calculation, on exactly integrable quantum chains \cite{alcaraz-rajabpour-prb14,alcaraz-rajabpour-prb15} show numerical 
evidence that these quantities, when computed on the ground state wave functions 
of critical chains expressed in special basis (conformal basis), exhibit some 
universal features. It is then interesting to compute these generalized 
mutual information for the extended $Z(Q)$ models introduced in this paper and 
test the
universal behavior for those critical non-integrable quantum chains. 

In order to define the generalized mutual informations let us split, as before, 
 the 
quantum chain $\cal C$ with $L$ sites in the subsystems $\cal A$ and $\cal B$ 
formed by  $\ell$ and ($L-\ell$) consecutive sites, respectively. We now 
consider the quantum chain in the normalized ground state, with wavefunction
$\ket{\Psi_{\cal C}}= \sum_{\{I_{\cal A},I_{\cal B}\}}
 a_{I_{\cal A},I_{\cal B}} \ket{I_{\cal A}} \otimes 
\ket{I_{\cal B}}$, 
where $\ket{I_{\cal A}}=\ket{i_1,i_2,\ldots,i_{\ell}}$ and 
 $\ket{I_{\cal B}}=\ket{i_{\ell+1},\ldots,i_{L}}$ are the local basis 
for the subsystems $\cal A$ and $\cal B$. The R\'enyi entropy for the 
entire system $\chi={\cal C}$ and the subsystems $\chi={\cal A}$ or 
$\chi={\cal B}$ are given by:
\beq \label{e23}
Sh_n(\chi) = \frac{1}{1-n} \sum_{\{I_{\chi}\}} \ln P_{I_{\chi}}^n, \quad 
\chi={\cal A}, {\cal B}, {\cal C},
\eeq
where for the entire system 
 $P_{I_{\cal C}} = |a_{I_{\cal A},I_{\cal B}}|^2$ and for the 
subsystems $\cal A$ and $\cal B$, 
 $P_{I_{\cal A}} = \sum_{I_{\cal B}} |a_{I_{\cal A},I_{\cal B}}|^2$
 and $P_{I_{\cal B}} = \sum_{I_{\cal A}} |a_{I_{\cal A},I_{\cal B}}|^2$,
respectively.
The R\'enyi mutual information is the shared information among the 
subsystems measured in terms of the R\'enyi entropy \rf{e23}, i. e., 
\beq \label{e24}
I_n(\ell,L-\ell) = Sh_n(\ell) + Sh_n(L-\ell) -Sh_n(L),
\eeq
where instead of denoting the subsystem, we denote their lattice sizes. At the 
limiting case $n\to 1$ the R\'enyi entropy and the R\'enyi mutual information reduces to 
the Shannon entropy and the Shannon mutual information, respectively. 
 
Previous calculations of $I_n(\ell,L-\ell)$ for the ground state wave functions of several exactly integrable chains show the same finite-size scaling 
function for arbitrary values of $n$:
\beq \label{e25}
I_n(\ell,L-\ell)= c_n \ln(\frac{L}{\pi}\sin(\frac{\ell \pi}{L})) + k,
\eeq
where $k$ is a $o(1)$ constant. As happens with the Shannon mutual 
information $I(\ell,L-\ell)$ this behavior is not general, it happens only 
when the ground state wavefunction is expressed on the special basis 
(conformal basis). The coefficients $c_n$ besides its $n$ dependence also 
depends on the conformal basis considered. Under certain plausible 
assumptions the large-$n$ behavior of $c_n$ is known analytically \cite{1742-5468-2014-5-P05010}. 
However in the general case the limiting case $n \to 1$ is singular, 
preventing a general analytical calculation of the Shannon mutual 
 information $I_1(\ell,L-\ell)=I(\ell,L-\ell)$. 

Our numerical analysis for the extended self-dual $Z(Q)$ models introduced 
in Sec.~II indicates the same universal finite-size scaling behavior 
shown in \rf{e25}. This confirmation was done for the values of the parameter 
$p$ that we believe the model share the universality class of critical 
behavior of the corresponding exactly integrable model ($p=0$).  For brevity 
we only show 
the results for the self-dual extended Ising models \rf{e3}. In 
 Fig.~15 and Fig.~16 the results  are for the quantum chain  with $L=30$ 
sites and the ground state wavefunction  spanned in the conformal bases where 
$\{\sigma_i^z\}$ or $\{\sigma_i^x\}$ are diagonal. In theses figures we show the 
coefficient $c_n$ obtained from the linear fit of \rf{e25}, by using 
all the sublattice sizes. We can see that in both basis, apart from some 
small  deviations, most probably due to the finite-size effects, the overall 
 behavior of $I_n(\ell,L-\ell)$ is 
the same for different values of $p$, indicating the universal behavior  of the 
models. It is clear from this figure  that the singular behavior as $n\to 1$, already 
known \cite{alcaraz-rajabpour-prb14} for the exactly integrable model ($p=0$), also happens for the extended 
Ising quantum chains with $p\neq 0$.
\begin{figure} [htb] \label{fig15}
\centering
\includegraphics[width=0.35\textwidth]{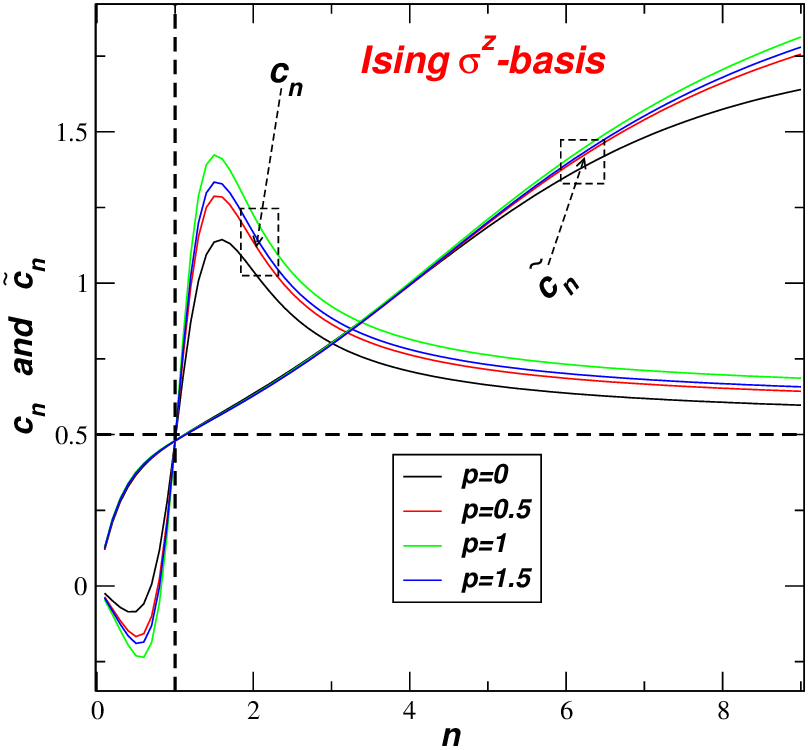}
\caption{ The generalized mutual informations for the ground state wavefunction of the 
extended Ising chain \rf{e3}, with $L=30$ sites. The coefficients $c_n$ and $\tilde{c}_n$ are 
obtained from the linear fit of \rf{e25} of the R\'enyi  mutual information 
$I_n(\ell,L-\ell)$ \rf{e23}-\rf{e24} and from the generalized mutual information 
$\tilde{I}_n(\ell,L-\ell)$, given by \rf{e26}, respectively.
The ground sates of the quantum chains are expressed in the $\{\sigma^z\}$ basis and 
the values of the parameter $p=0,0.5,1$ and 1.5 . } 
\end{figure}
\begin{figure} [htb] \label{fig16}
\centering
\includegraphics[width=0.35\textwidth]{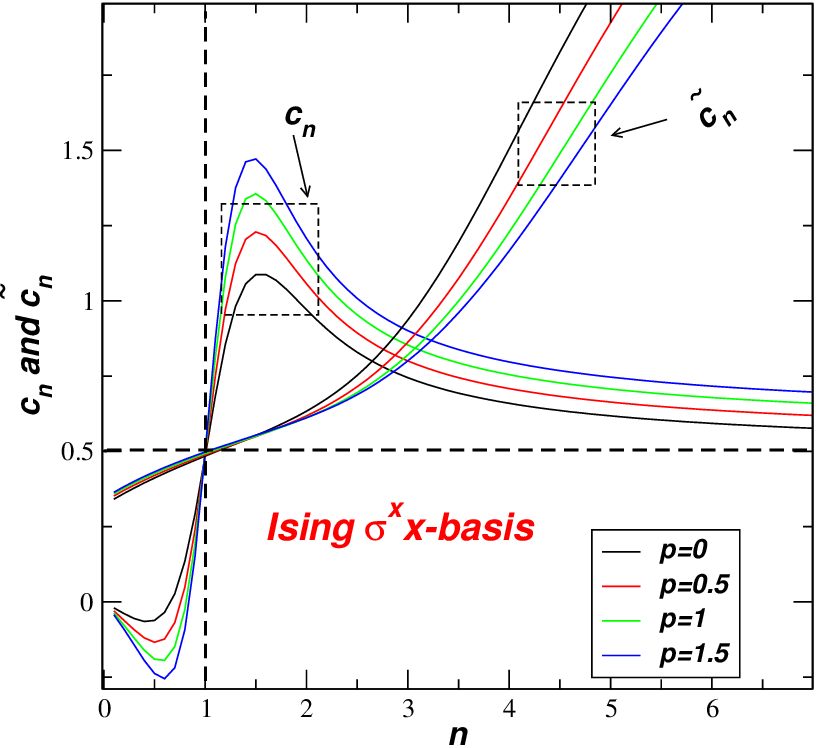}
\caption{ 
Same as Fig.~15, but with the ground state wave function spanned in the $\{\sigma^x\}$ basis.}
\end{figure}

Another interesting generalization of the Shannon mutual information, instead 
of being based in the R\'enyi entropy is based in the R\'enyi 
divergence \cite{Principe:2010:ITL:1855180}. Differently from the R\'enyi mutual information this 
generalized mutual information is always a positive function and is a more 
appropriate measure,  
from the point of view of information theory, of the   shared information among subsystems. 
 Using the notations in \rf{e23} this generalized 
mutual information is defined by:
\beq  \label{e26}
\tilde{I}_n(\ell,L-\ell) 
= \frac{1}{n-1}\ln \left( 
\sum_{\{I_{\cal A},I_{\cal B}\}}
\frac{
P_{I_{\cal A},I_{\cal B}}^n}
{P_{I_{\cal A}}^{n-1}
P_{I_{\cal B}}^{n-1}}.
\right)
\eeq
Like $I_n(\ell,L-\ell)$ this quantity, in the limiting case  $n\to 1$, gives the Shannon mutual 
information. 
This quantity was measured for several exactly integrable quantum chains \cite{alcaraz-rajabpour-prb15}.
 It shows the same universal finite-size scaling function given in \rf{e25} for
 $n\lesssim 2$ (we denote the linear coefficient as $\tilde{c}_n$). We measured this quantity for the extended $Z(Q)$ models 
introduced in Sec.~2. The results for the extended Ising quantum chain are 
shown in Figs.~15 and Fig.~16 for the ground state 
wavefunction expressed in the $\{\sigma^z\}$- and $\{\sigma^x\}$-basis, respectively.
 Again for $0<n<2$ we clearly see in both basis the independence of the curves 
with the parameter $p$ of the non-integrable quantum chain.  Actually 
the agreement of this behavior for several values of $p$ is even better 
as compared with the case of the R\'enyi mutual information, this indicates 
that the finite-size scaling corrections in $\tilde{I}_n(\ell,L-\ell)$ are 
smaller than  the ones in $I_n(\ell,L-\ell)$. It is also clearly shown that 
the limiting case $n\to 1$ is regular for all values of $p$, differently 
from the case of the R\'enyi mutual information. This imply that 
$\tilde{I}_n(\ell,L-\ell )$, as compared with $I_n(\ell,L-\ell)$ is  
 a more suitable quantity for an analytical 
approach towards the proof of the conjecture \rf{e2}. 

\section{Conclusions}

In this paper we made an extensive test of the conjecture \rf{e2} for the Shannon mutual 
information $I(\ell,L-\ell)$ of conformally invariant quantum critical chains at their ground 
states. In general the Shannon mutual information depends on the particular basis where 
the wavefunction is spanned. According to the conjecture \rf{e2} the finite-size scaling 
function of $I(\ell,L-\ell)$ give us an interesting tool for calculating the central 
charge $c$, if the ground state is spanned in the conformal basis. These basis 
corresponds, in the underlying Euclidean CFT, to the boundary condition in the time 
direction that do no destroy the conformal invariance of the CFT. 

This paper provide us with the first extensive numerical check of the universal character 
of \rf{e2}. The previous tests of \rf{e2} were done only for exactly integrable quantum 
chains, and since there is no analytical proof of \rf{e2} it is important to verify if its 
validity is not connected to the exact integrability of the critical quantum chains 
tested previously.

In order to produce tests for non-integrable models we introduced new families of self-dual 
quantum chains with nonlocal $Z(Q)$ symmetries. Due to their self-duality their critical 
points are exactly known. All these non-integrable quantum chains contains next-nearest 
neighbor coupling constants $p$. Our  numerical analysis concentrated in two special families 
of models. The first family is the generalization of the $Z(Q)$ parafermionic models ($Q=2-8$), 
and the second one is the generalization of the $Z(Q)$ clock models ($Q=5-8$). The first 
family at $p=0$ reduces to the exactly integrable parafermionic quantum 
chains with central charge $c=\frac{1}{2},\frac{4}{10},1,\frac{8}{7},\frac{5}{4},\frac{4}{3},
\frac{7}{5}$, for $Q=2-8$, respectively. The second family reduces at $p=0$ to non-integrable 
quantum chains in the Beresinzkii-Kosterlitz Thouless universality, whose underlying CFT is 
expected to have a central charge $c=1$ for $Q \geq 5$. Exploring the consequences 
of conformal invariance, our numerical studies of the low-lying energies of these quantum 
chains, at finite lattice sizes, indicate that at least for a finite range of the couplings 
$0 \leq p\leq p_c$ the models share the same universal critical behavior, and consequently are 
ruled by the same CFT.

The last observation make these introduced quantum chains even more interesting, since as we 
change continuously the parameter $p$ they give a critical line with a fixed value of the 
central charge. In particular the extended parafermionic quantum chains for $Q\geq 5$ give us 
critical lines ruled by an underlying $Z(Q)$ parafermionic CFT with $c> 1$.

The extensive calculations of the Shannon mutual information $I(\ell,L-\ell)$ of the 
ground state wavefunctions of all these quantum chains indicate the validity of the 
conjecture \rf{e2} for general critical and conformally invariant quantum chains, irrespective 
of being exactly integrable or not. 

It is important to mention that St\'ephan \cite{stephan-mutual-ising} presented 
a contradictory 
prediction for the critical Ising quantum chain. In \cite{stephan-mutual-ising},
 by exploring 
the free-fermionic nature of the model, $I(\ell,L-\ell)$ was calculated numerically up
 to lattice 
sizes $L=36$, and the results indicate that the pre-factor in \rf{e2} instead of 
being the 
central charge $c=0.5$, is the close, but distinct number $c=0.4801629(2)$. This
 would imply 
that the conjecture \rf{e2} is not valid and the pre-factor is a universal unknown 
number whose value is close to the central charge, at least for the Ising case. 
All the 
numerical results we have obtained so far for the several quantum chains does 
not have enough 
precision to discard the possibility that for all the critical chains the pre-factor in the 
 conjecture \rf{e2} could not be the central charge $c$, but a number close 
to it. The 
single exact analytical exact calculation we have is for the set of coupled 
harmonic oscillators 
that gives in this case the central charge value $c=1$ \cite{alcaraz-rajabpour-prl13}. The result 
in \cite{stephan-mutual-ising} was obtained by assuming that the finite-size 
corrections of 
$I(\ell,L-\ell)$ are given by the power series $\sum_{p=0}^5 \alpha_p/ \ell^{p}$, being 
the fitting 
quite stable indicating no presence of logarithmic corrections, like $\frac{ln{\ell}
}{\ell}$ terms.

As is well known in order to have a controlled prediction of quantities in the bulk limit, based 
on finite-size lattice estimators we should know the functional dependence of the finite-size 
corrections with the lattice size. Unfortunately this is not the case for 
$I(\ell,L-\ell)$. This is 
an essential point. $I(\ell,L-\ell)$ is calculated by combining the probabilities $p_
{\{x\}}$ of the 
configuration $\{x\}$ in the subsystem of size $\ell$. The probabilities for special 
configurations of the Ising quantum chain can be calculated for quite large 
lattices $L\sim 1000$.
 The results for $\epsilon(\{x\}) = -\ln p_{\{x\}}$, also called as the formation
 probabilities, 
shows that for special commensurable configurations $\{x\}$, like the emptiness 
formation 
probability and generalizations (see appendix of \cite{PhysRevB.93.125139}), indicate that 
correction terms $\frac{\ln{\ell}}{\ell}$ are always present. If as a result of the combinations 
of the several probabilities in $I(\ell,L-\ell)$ these logarithmic corrections are canceled then 
the prediction of St\'ephan \cite{stephan-mutual-ising} is correct and the conjecture has to be 
modified. On the other hand if still these corrections are present in 
$I(\ell,L-\ell)$, then  
we should consider lattice sizes  or order $L\sim 1000$ to discard or to confirm the 
conjecture \rf{e2}. This is indeed a quite interesting point to be settled in the future. It is 
a challenge either to derive analytically $I(\ell,L-\ell)$ or at least to derive the 
behavior of 
the finite-size corrections.


There exist two extensions of the Shannon mutual information, namely The R\'enyi mutual 
information $I_n(\ell,L-\ell)$ and the generalized mutual information $\tilde{I}_n(\ell,L-\ell)$, based on the R\'enyi divergence.
These quantities were calculated previously for several exactly integrable quantum chains in their 
ground state. As the Shannon mutual information they also show some universal features 
whenever the ground state wavefunction is spanned in a  conformal basis. We calculate the 
generalizations $I_n(\ell,L-\ell)$ and $\tilde{I}_n(\ell,L-\ell)$ for the non-integrable 
models introduced in this paper. Our results indicate that the universal features previously 
observed \cite{alcaraz-rajabpour-prb14,alcaraz-rajabpour-prb15} does not depend if the quantum chain is exactly integrable or not. It  is 
important to mention that, as happen for the exactly integrable cases \cite{alcaraz-rajabpour-prb15}, 
$\tilde{I}_n(\ell,L-\ell)$ in general does not have a divergence as $n\to 1$, differently from 
the generalization $I_n(\ell,L-\ell)$. Since this divergence destroy the analytical 
continuation $n\to 1$, the quantity $\tilde{I}(\ell,L-\ell)$ seems to be more appropriate 
for an analytical derivation for the conjecture \rf{e2} for the Shannon mutual information 
$\tilde{I}_1(\ell,L-\ell)=I(\ell,L-\ell)$.

\textit{Acknowledgments}
This work was supported in part by FAPESP and CNPq (Brazilian agencies). We thank M. A. Rajabpour and J. A. Hoyos for useful discussions.

\bibliographystyle{apsrev4-1}
\bibliography{/home/alcaraz/tex/bibliography/referencias-alcaraz}

\end{document}